# Efficient reconfigurable regions management method for adaptive and dynamic FPGA based systems

Marwa Hannachi[1, 2,*], Abdesslam B.Abdelali[1], Hassan Rabah[2], Abdellatif Mtibaa[1]

[1] *Laboratory of Electronics and Microelectronics, University of Monastir, Tunisia*
[2] *Institut Jean Lamour (IJL) UMR7198, University of Lorraine, France*
*Corresponding author: marwa.hannachi@univ-lorraine.fr

**Abstract**

Adaptive systems based on field programmable gate array (FPGA) architectures can greatly benefit from the high degree of flexibility offered by dynamic partial reconfiguration (DPR). By using this technique, hardware tasks can be loaded and reloaded on demand depending on the system requirements. In this paper, we propose to use the DPR for dynamic and adaptive implementation of a video cut detection application based on the MPEG-7 color structure descriptor (CSD). In the proposed implementation, different scenarios have been tested. Depending on the application and the system requirements, the CSD module can be loaded at any time with variable module size (corresponding to different version of the CSD) and allocated in different possible reconfigurable regions. Such implementation entails many problems related to communication, relocation and reconfigurable region management. We will demonstrate how we have made this implementation successful through the use of an appropriate design method. This method was proposed to support the management of variable-size hardware tasks on DPR FPGAs based adaptive systems. It permits to efficiently handle the reconfigurable area and to relocate the reconfigurable modules in different possible regions. The implementation results for the considered application show an important optimization in terms of configuration time (until 66 %) and memory storage (until 87 %) and an efficient hardware resources utilization rate (until 90%).

## 1. Introduction

DPR technique offers the ability to modify the configuration (circuit structure) of an FPGA part, while the rest of the device continues working without interruption (Hong *et al*. 2014; Lysaght *et al*. 2006). In DPR systems, the architecture is partitioned into static and partial reconfigurable regions (PRR). In order to enable an application to change the PRR structure, a set of configurations, known as partial reconfigurable modules (PRMs), is required. The latter will be substituted and placed in a predefined PRR on the FPGA.

FPGA-based DPR is greatly appropriate choice for systems needing a high degree of flexibility. This promising technique can be applied to deal with: lack of available resources, reduction of energy consumption, system adaptability and system reliability. The following situations can benefit from DPR:
- Available resources are smaller than system's resource requirement.
- Necessity of reducing system energy consumption.
- The application is composed of multiple independent and time exclusive tasks (can be implemented sequentially by sharing the same hardware resources).
- Some hardware resources are not always needed.
- System used for multiple application contexts
- System supporting components upgrade (with improved or updated versions).
- System functions have to be changed depending on the application environment.

One of the key challenges in DPR FPGA based architectures design is the reconfiguration and the PRR management: relocation, communication, etc. In this paper, we propose a dynamic and adaptive implementation of a video analysis application, while fixing the problems related to communication, relocation and reconfigurable region management. We propose a dynamic implementation of a cut detection application based on the MPEG-7 CSD. Depending on the application and the system requirements, the CSD module can be loaded at any time with variable module size (corresponding to different quantization levels) and relocated in different possible reconfigurable regions. The use of the DPR technique with partial bitstream relocationpresents many benefits in our application such as: the possibility of relocating the bitstream generated for a specific PRR on different other ones, eliminating the need to multiple PRB for the same hardware task, reducing the amount of memory used to store partial bitstreams and allowing runtime bitstreams placement. However implementation of this technique entails many problems related to communication and area management (related to variable size modules). In fact, the same communication constraints must be maintained between DRRs for the different related PRMs. Also, for variable size modules occupying the same region, an appropriate management strategy is necessary to increase the area use and the configuration time efficiency. To address these problems, a method supporting variable-sized hardware tasks will be proposed and applied for the considered application.

The rest of this paper is organized as follows: Section 2 is dedicated to the application context description and the problem formulation. The DPR application scenario and the different problems involved will be also addressed in this section. In section 3, we will describe the considered application of video cut detection based on the CSD and the reconfigurable hardware modules design. The proposed method will be detailed in Section 4. Section 5 will be dedicated to describe the experimental results obtained during the CSD based cut detector approach.

## 2. Overview of the proposed application scenario and problem formulation

Modern multimedia systems are characterized by a rich set of services from which the user can easily make a choice to switch between the different given ones. It can also add new services or substitute the existing ones according to its preference. In this case, quality of Services (QoS) is of a big importance; it is used as a distinct factor between similar services (Aljazzaf, 2015). To support such functionalities and the high requirements in terms of multi-processing capacity, portability and consumption, the use of adequate hardware systems with high level of adaptability becomes necessary. In this context, the DPR utilization can be a very interesting solution. An appropriate application of this technique permits to support complex scenarios of adaptive and dynamic multimedia treatments implementation.

The considered application scenario in this work is illustrated in Figure 1. Given a set of multimedia treatments, an FPGA partitioned into dynamically reconfigurable regions, and relative to the choice made by the user, different modules treatments will be loaded into the FPGA. These modules can be allocated in different reconfigurable regions, depending on the system configuration at the moment of the new service selection. Also, adaptive version of different treatment modules can be loaded depending on the application environment and system requirements. These modules are given with different service qualities and physical sizes.

As an example of such complex reconfiguration management, we propose an adaptive and dynamic implementation of a video cut detection system based on the MPEG-7 CSDin this paper. As illustrated in Figure 1, the CSD exists with different quantization levels, and it can be allocated in different reconfigurable regions with a variable module size. In fact, as demonstrated in Ben Abdelali *et al*.(2014), the CSD can be implemented for different quantization levels as: 16, 32, 64, etc. It was demonstrated that the use of a low number of quantization levels can significantly reduce the algorithm complexity and the occupied hardware resources, while preserving a satisfactory level of accuracy in terms of cutdetection rate. This is useful, when the system is constrained by occupied hardware resources or consumption. In this case, a lower complexity version of the CSD can be loaded to respond to the system requirements.

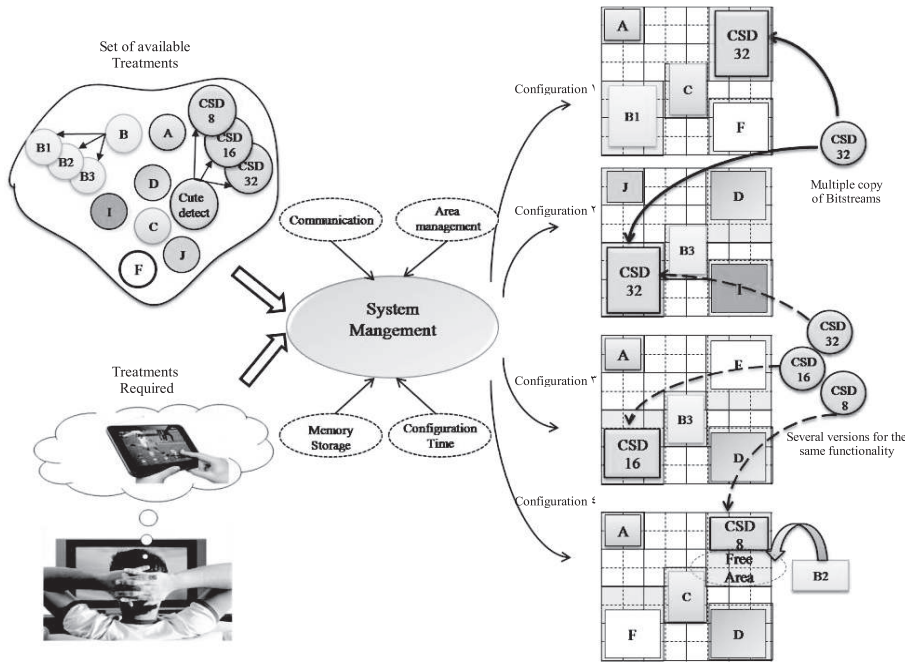

**Fig.1**.Proposed application

Such complex implementation scenarios of the CSD evoke many problems that have to be handled. The latter can be summarized in the following points:

-Relocation: In the Xilinx PR design flow, PRRs have to be designed and fixed before bitstreams generation. A set of PRMs will be assigned to each PRR in the implementation design flow. The same PRM can be assigned to different PRR. Thus, for each PRM different PRBs (PRB) will be generated for the different associated PRRs. In this case, multiple PRBs must be created for the same module. A large memory is then necessary to store the different required PRBs. An appropriate relocation technique can be used to deal with this problem. The solution here is to generate only one bitstream for the initial location of each PRM, and to adapt it for the different other locations.

Reconfigurable area management: In reconfigurable systems, the size of a reconfigurable region must be fixed at the beginning of the design flow. When variable-size hardware tasks are assigned to a given reconfigurable region, it must be sufficiently large to be able to take the biggest one. As a consequence, the biggest task will employ the whole available area of the reconfigurable region.However, tasks with smaller sizes may have a muchlower rate of area use, which leads to a non-optimized employment of the hardware resources.To deal with this problem, a size adaptation method of PRR to hardware tasks size will be applied for a better management of the PRR. As a solution, the reconfigurable region will be partitioned into a number of small regions, called partitions, as proposed in Marques *et al*. (2014).The partitioning is done in a way to ensure a high occupation rate and a low reconfiguration time.

Communication: Allowing dynamic insertion and removal of PRMs for different locations entails a critical problem related to the communication between the different reconfigurable and static modules. To ensureefficient relocation for such reconfigurable tasks, the communication problem must be taken into account. The same communication primitives (partition pin and routing paths) must be maintained for all reconfigurable partitions in all reconfigurable regions corresponding to a given PRM.

## 3. Presentation of the implemented video cut detector system

In this section, we will describe the developed video cut detection system based on the MPEG-7 CSD. The principle of cut detection application using the CSD, and the developed hardware architectures will be presented. Synthesis results of these architectures will be given for different CSD application modes relative to the input frame data quantification levels.

3.1 Video cut detection based on the CSD

Video cut detection techniques permit to detect an abrupt change in a video sequence (Cernekova & Pitas 2006); Krulikovská & Polec (2012). These techniques are generally based on visual dissimilarity measurement between consecutive frames relative to a given low level feature (histogram, color distribution, texture, etc.) (Ben Abdelali *et al*. 2014; Yu & Srinath, 2001).The obtained measure should present a significant change in order to determine the existence of a CUT. In our application, the CSD was applied to measure the difference between consecutive frames.

The principle of cut detection based on the CSD is illustrated in Figure 2. The CSD is extracted for each frame and a distance calculation is performed to measure the variation of color structure histogram between each two consecutive frames. A cut is detected, if the obtained distance exceeds a given threshold.

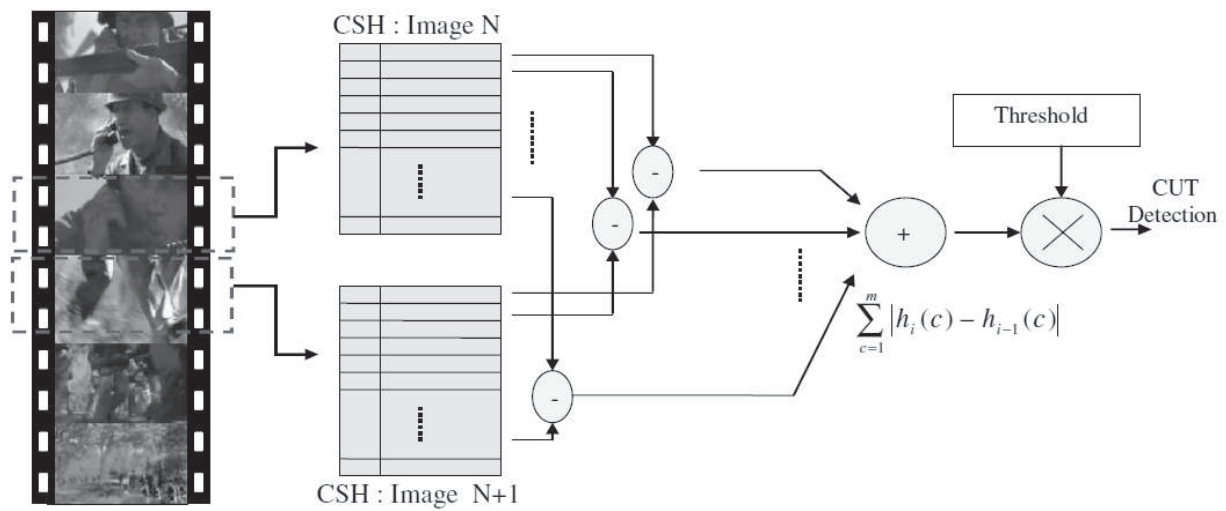

**Fig. 2.** Cut detection principle

The CSD permits to express the local color structure in an image (Manjunath *et al.*, 2001; Vertan *et al.*, 2011). It aims to provide a more accurate description by identifying local color distribution. The CSD is identical to a color histogram in form, but it is different in terms of semantics. The color structure histogram is computed by visiting all image locations using an 8x8 structuring element, observing which colors are presented in it, and then updating color bins (Ben Abdelali & Mtibaa, 2005). The incrimination of the bins is commanded by the presence of the corresponding colors; not by the count of each present color pixels. Therefore, in any given position of the structuring element, the increase can be either 0 or 1. The color structure histogram is represented by 1D array and it can be denoted by:

$$h = (h(1), h(2), ..., h(m))$$

With h(i), i = 1,2, …,m, represents the value of the ith histogram bin, it gives the number of structuring elements in the image containing one or more pixels with color "Ci". "m" indicates the number of color bins, which is equal to the quantization levels number.

After the CSD extraction, the visual discontinuity between consecutive frames is calculated using the difference of color structure histograms. We employ the Manhattan distance (L1 distance) to define frame distance. We have two frames represented by two "m" dimensional vectors (hi(1) hi(2), . . . , hi(m)), (hi-1(1), hi-1(2), . . . , hi-1(m)), respectively. The Manhattan distance is defined as:

$$d(F_i, F_{i-1}) = \sum_{c=1}^{m} |h_i(c) - h_{i-1}(c)| \quad (1)$$

where hi: the color structure histogram of frame i,

hi-1: the color structure histogram of frame i-1,

m: the number of color quantization levels.

If the obtained distance exceeds a given threshold, a shot change will be reported and one of the first frames will be considered as the representative key frame of the current shot.

Hardware architecture of the cut detection system based on the CSD

The hardware architecture of the cut detection system based on the CSD is illustrated in Figure 3. This architecture is composed of two main parts: the CSD module and the "distance calculation" module. The latter permits to calculate the histogram distance between consecutive frames. Then, it compares the obtained distance to a given threshold. The CSD block output is the histogram bins, whose number depends on the number of quantization levels. When a frame is processed, the new CSD output vector will be stored in the register set 1 (dedicated to store the CSD of frame i), and the old one previously stored in the same register set will be transferred to register set 2 (dedicated to store the CSD of frame i - 1). After that, the distance calculation module starts a new distance measurement between the current consecutive frames. One adder, one subtractor and one register are used to calculate and to store the result. In each time, a comparison is made between the obtained distance (d) and the given threshold ($\alpha$). If d >$\alpha$, a scene change is detected between frames 'i' and 'i-1' and the out signal 'Detect_EN' will be activated.

The color detection block allows the detection of the existent colors in the (8×8) structuring element. It is composed of n "color detector" blocks (n = number of colors). Each color detector is hierarchically applied line by line on the 8X8 window. We obtain 8 "line detector" blocks for each color detector. Each line detector is used to detect the existence of the color in the corresponding line. It has as an input, the 8 registers values of the corresponding line: Li-R1 to Li-R8; where Li-Rj {i=1,…, 8, j=1,…,8} represents the register storing the (i, j) point of the 8x8 window. Each register output will be compared to the desired color; if at least one pixel has this color, the block output will be activated. If at least one of the 8 line detector blocks output (of a given color) is at '1' logic, the color existence will be reported and the corresponding histogram bin will be incremented. The color detection block has n bits as output; each one indicates the presence of the corresponding color. These outputs are related to the histogram update block inputs.

The histogram update block is composed of n counters (n = number of colors). Each counter will be incremented, if its input is at '1' logic. The counter width depends on the number of possible structuring element positions (NP) for the considered image size. The number of positions (NP) is given by:

$$NP = (height - sw + 1) \times (width - sw + 1) \quad (2)$$

where "height", "width", and "sw" represent respectively the image height, the image width and the structuring element width. In our case sw=8, height =640, width = 480 and NP = 299409. The counters width (Cn) is calculated such that:

$$2^{Cn} \geq NP \quad (3)$$

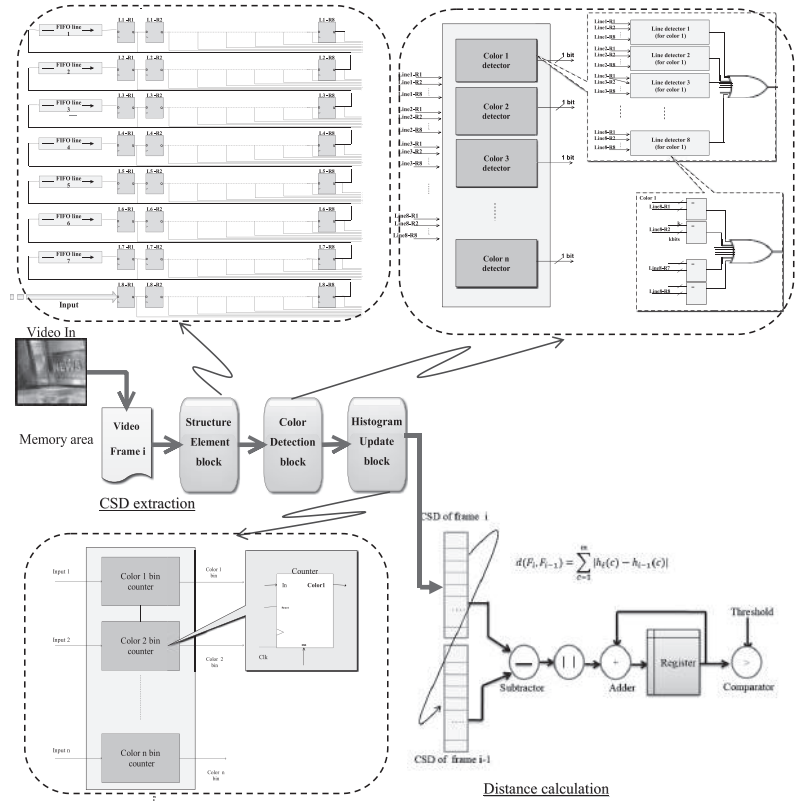

**Fig. 3**. System description

The cut detection system was synthesized for an XC5VSX50T FPGA using Xilinx ISE tool. The proposed architecture was implemented for different number of quantization levels (8, 16, and 32).The detailed synthesis results in terms of number of occupied slices, LUTs, registers, RAM Blocks and maximal frequency depending on the number of quantization levels (n) are given in Table 1. The obtained results show that the use of a lower quantization levels number has a significant effect on the hardware resources occupation. The number of occupied slices varies from 1445 for 32 quantization levels to 499 for 8 quantization levels.

**Table 1.** Synthesis results for different quantization levels

| Number of colors(n) | Slices | Slice LUTs | Slice Registers | BRAM | Frequency (MHZ) |
|---|---|---|---|---|---|
| 8 | 499 | 1304 | 1373 | 4 | 215.983 |
| 16 | 1008 | 2498 | 2173 | 4 | 201.045 |
| 32 | 1445 | 4318 | 3710 | 4 | 199.800 |

## 4. The proposed method for variable size hardware tasks relocation

In this paragraph we will discuss the proposed method to handle the issues defined in section 2. This method is based on PRB relocation techniques with a main focus on communication and area management. Our objective is to generate only one bitstream for each CSD version independently of the different locations that will be occupied by it. The generated bitstream can be manipulated in the reconfiguration phase to be adapted to the target location. In the implementation phase, the communication between different static and reconfigurable regions and the hardware resources distribution must be rigorously taken into consideration (Hannachi *et al*., 2014). An efficient management of the reconfigurable areas is also an important objective of the proposed method. The aim here is to ensure an efficient area occupation permitting to minimize the reconfiguration time relative to the reconfigurable modules sizes.

4.1 State of the art and principle of the proposed method.

The PRB relocation (PBR) technique consists in manipulating a bitstream in order to adapt it to different locations than the one for which it was generated (Flynn *et al*., 2009; Sudarsanam *et al*., 2010). A bitstream is generated only for the initial location of each module, and the relocation technique will be used to change the physical bitstream location on the FPGA. This technique permits to eliminate the need for multiple PRBs for the same task and to reduce the amount of memory used to store them. However many considerations must be taken into account in the design phase. The use of homogenous reconfigurable regions (identical hardware resources in number and disposal) presents, for example, a main constraint to be taken into consideration to successfully make the relocation phase (Becker *et al*., 2007).

Although, several works have been carried out in the bitstream relocation subject, different important aspects were ignored or not properly taken into consideration. In papers Corbetta *et al*. (2009), Horta *et al*. (2001), Kalte *et al*. (2005), Sudarsanam *et al*. (2010) and Touiza *et al*.(2013), various methods and tools were developed to support PBR. However, the works presented in these papers focused mainly on the location information extraction and bitstream manipulation aspect, without details about the implementation phase. In Drahonovsky *et al*.(2013) and Ichinomiya *et al*. (2012), the authors propose design flows, which are based on the proxy logic position and routing constraints modification. These flows require the creation of reconfigurable regions with identical partition pins position for the various locations of the PRM in the FPGA device. In all these papers, the size of each reconfigurable region is fixed during the design phase, relative to the biggest PRM, which leads to an inefficient area occupation, when dealing with variable sized IP cores.

In our relocation method, all these aspects will be taken into consideration in the same design flow: hardware resources distribution, communication resources organization and area management for variable sized hardware modules. For area management we will adopt the method proposed in Marques *et al*. (2014), which will be combined with the other implementation steps with the aim to assure an efficient hardware resources occupation and reconfiguration time.

As shown in Figure 4, in a classic relocation method (Figure 4 b), the whole PRR is reserved either if the module size is lower than its area, which leads to a high rate of wasted FPGA area. Also, for this method, the reconfiguration time is the same for all modules independently of their size. In our approach (Figure 4. c), a better adaptation of PRR is proposed. Here, the main idea consists on virtually partitioning the PRR in a set of partitions adapted to the corresponding PRMs sizes. Different possible locations will be also defined for each PRM in the PRR. A design will be generated for each possible location of the different modules. For example, the PRMs assigned to the PRR1 are M1, M2 and M3 (Figure 4 a). This region is partitioned into three partitions (M3- M3-PRP1, M3-PRP2 and M3-PRP3) for M3, two partitions for M2 and only one partition for M1, which represents the biggest module. By applying this method, an important gain in terms of reconfiguration time will be obtained, since we will reconfigure only the partition corresponding to the current PRM and not the whole PRR. Also more intelligent reconfiguration management can be applied with this PRR partitioning. For example, when we use M3-PRP1 or M3-PRP2 for M3, the M2 module can be preloaded in the remaining area with reconfiguration time masking, and this permit to obtain more efficiency in terms of reconfiguration time.

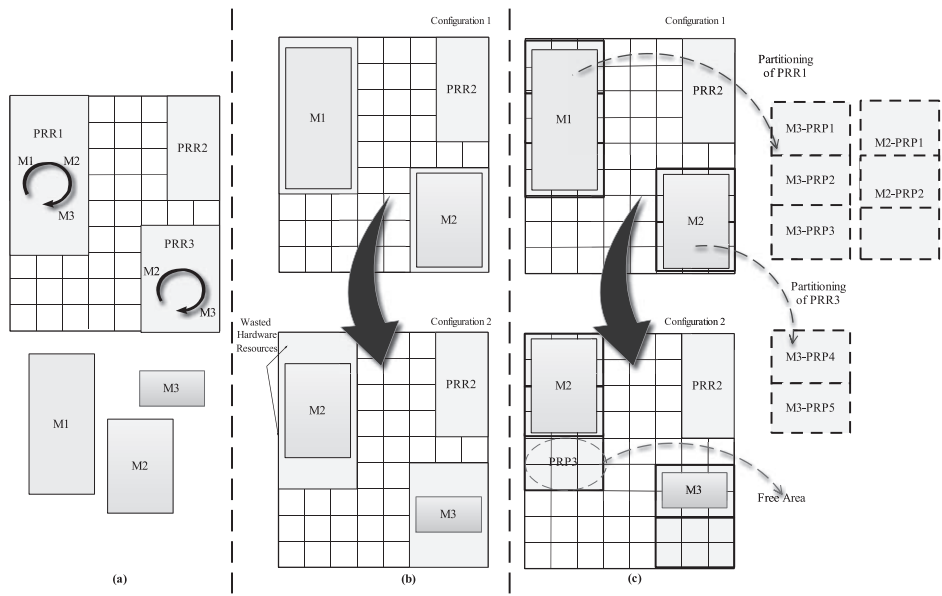

**Fig. 4.** Principle of PBR combined to PRRs partitioning: (a)PRRs definition,
(b) Classical relocation, (c) Proposed relocation

In Table 2, a comparison with existing design flows is presented. The most important features of the following methods are listed: Xilinx DPR design flow (Xilinx, 2012b); RapidSmith 2(Haroldsen *et al*., 2015); OpenPR (Sohanghpurwala *et al*., 2011); RecoBus-builder (Koch *et al*. 2008); GoAhead (Beckhoff *el al*., 2012) and CoPR (Vipin & Fahmy, 2015). The last column of Table 2 gives features of the proposed method. As we utilize the latest Xilinx tools, the Partition Pin is employed as the communication interface between reconfigurable and static regions. The proposed method mainly offers the possibility of partitioning a PRR into several partitions. This feature, which is not supported by the other design flows, allows adaptation of the PRR size to the hardware IP core and a better management of PRRs. The main drawback of the classic implementation methods is the high inefficiency in terms of resources utilization. In this case, unused hardware resources of PRR are wasted and cannot be used for other hardware IP cores.

As another important feature, in our design flow, we address the physical level aspect of the relocation technique, which is not supported by the classic Xilinx design flow and partially addressed by some other ones.

**Table 2.** Comparison with existing methods

| | Xilinx | RapidSmith2 | OpenPR | RecoBus | GoAhead | CoPr | Proposed method |
|---|---|---|---|---|---|---|---|
| **Design Tools** | Xilinx tools | RapidSmith 2 | Xilinx tools | RecoBus | GoAhead | Xilinx tools | Xilinx tools |
| **Communication interface** | Partition Pin | Bus macros | Bus macros | Bus macros | Partition Pin | Partition Pin | Partition Pin |
| **Communication wrapper** | Direct wire, Bus Streaming | Direct wire | Direct wire | Direct wire | Direct wire, Bus Streaming | Direct wire, Bus Streaming | Direct wire, Bus Streaming |
| **Partitioning** | No | No | No | No | No | No | Yes |
| **Module relocation** | No | No | Yes | Yes | Yes | No | Yes |

### 4.2 Steps of the proposed method

The FPGA architecture is divided into two parts: static and dynamic. The static part contains hardware tasks that will not be modified during the application execution and the dynamic part contains the PRMs. The PRBs relative to the latter are generated only for the initial location of each PRM and then manipulated by modifying the fame address register (FAR) to relocate the hardware modules from a given location to others on the FPGA. To successfully perform this technique, the following constraints have to be taken into consideration in the implementation phase: compatibility of the reconfigurable partitions, same location and characteristics (BEL and LOC) of the partitions pins for all reconfigurable partitions and same routing paths between all partitions.

Given an FPGA with a number of dynamic PRRs and a set of modules to be allocated in these regions, the proposed method will be applied throughout the following steps:

Step 1: Partitioning of PRR:

In this step a partitioning of the PRRs and the different possible locations of each PRM are defined. Also the conformity of the PRR I/O ports is checked. The partitioning is performed for each module relative to its requirements in terms of hardware resources and by taking into consideration the physical structure of the used FPGA. This device is organized in frames, which are the smallest reconfigurable units. Three types of frames can be distinguished relative to the FPGA resources type: CLB frames, BRAM frames, and DSP frames. For maximum efficiency, the partitions relative to a given PRM should be chosen in the manner to maximize the rate of area utilization in terms of frames $Ra_{(frames)}$. $Ra_{(frames)}$ is calculated as follows:

$$Ra_{(frames)} = \frac{\sum_i W_i}{\sum_i N_i}, \quad i \in CLB, DSP, BRAM \quad (4)$$

where $W_i$ is the frames number of type i required to implement a given module and $N_i$ is the frames number of type i relative to the reserved partitions to allocate a given module.

Step 2: Create initial design:

In the initial design, each PRM is assigned to an initial location (partition) and the rest partitions are left as blank. A design will be generated for each module with its different possible locations and only a bitstream will be created for the first location. To be able to relocate the generated PRB from one partition to another, the resources in both partitions must be similar in terms of both resources type and position in the PRR.

Step 3: Extracting information related to partition pins:

Partition pins are used to ensure communication between the static and reconfigurable parts. Partition pins employed one look-up table (LUT) per bit defined in the reconfigurable partition side (Xilinx, 2012b). Partition pins placement in the PRR and the routing path of the related signals in the static part are automatically generated. While the partition pins positions are easy to identify, it is hard to extract the routing paths. In fact, the placement and routing of partition pins differ from a PRP to another and from a design to another. However, routing paths must be imposed in the UCF file withthe aim to keep the same crossing wires routing for all PRPs.

In order to deal with this problem, we propose to transform proxy logic into an equivalent bus macro, by creating a new PRR in front of the partition or by defining the whole static part as reconfigurable. Thus, an automatic insertion of LUT based pins in this region (static) can be made to build a macro bus as shown in Figure 5. In our design, the static part is defined as a PRR, to be able to create the new partition pins for the macro bus.

The placement of proxy logic (partition pin) is characterized by PIN and LOC (XY coordinate) constraints (Sudarsanam *et al*., 2010). The location and the type of basic element logic (BEL) for each partition pin in the device, corresponding to the initial design, are extracted. The UCF file is used to lock the placement of the communication interfaces for the other PRRs.

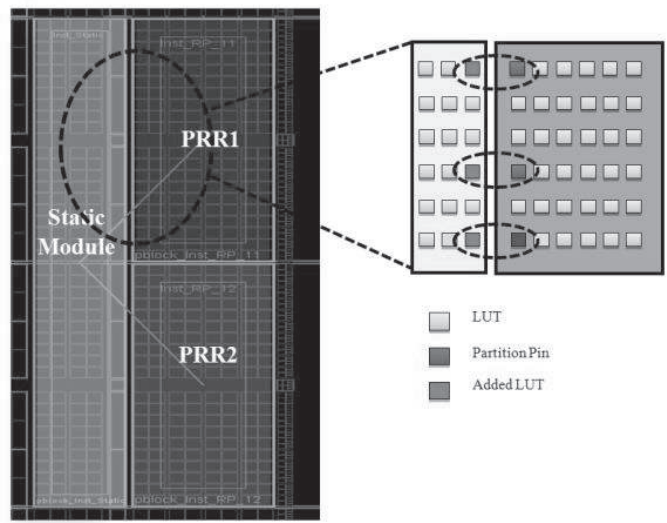

**Fig. 5.** Proxy logic placement

Step 4: Modifying the partition pins: This step consists of modifying the LOC and BEL of the partition pins. The same parameters of proxy logic (BEL and LOC) must be applied for the different partitions allocating the same PRM. The previously extracted parameters for the initial design are applied for the new reconfigurable partitions. The location of partition pins (X and Y coordinates) in the new partition must be modified, such that the pins take the same location in the reconfigurable partition as shown in Figure 6. a. These modifications are introduced as constraints in the UCF file and a new run of the design implementation is performed.

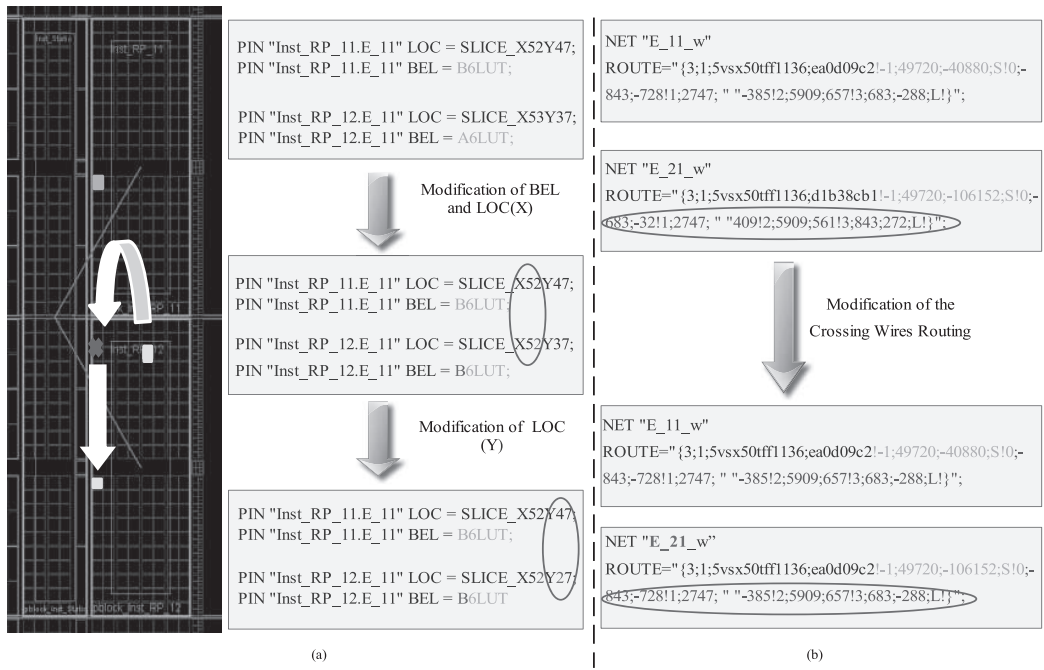

**Fig. 6.** Constraints modification: a. Partition Pins, b. crossing wires routing

Step 5: Extracting and modifying the crossing wires routing: After the second design implementation run with the partition pins modification in the previous step, all pins on the different reconfigurable partitions are placed in their right place. In this phase, it is necessary to extract the crossing wires routing between each two partition pins (macro bus between the partition pins of the reconfigurable partition and the added proxy logic in the static part). The same routing paths will be applied for all partitions in order to preserve them and to avoid their inclusion with any routing path of the static part. To apply the extracted routing for the different reconfigurable partitions, it is necessary to copy the crossing wires routing for all reconfigurable partitions and just change the address by the appropriate source address. The absolute address of signal source is between "! - 1" and "S! 0" as shown in Figure 6.b. In the same way as for partition pins, the routing paths are also introduced as constraints in the UCF file.

Step 6: Generating PRBs: A new design is created with the UCF file containing partition pins information (LOC and BEL) and the crossing wires routing. A design is created for each partitioning defined in step 1. This is done using the same UCF file, which permits to preserve the partition pins placement and the routing paths as for the initial design. Bitstreams for all PRMs are generated for the initial locations (partitions).

Step 7: Modifying bitstreams (relocation): This phase consists of manipulating the generated bitstreams in the aim to modify the corresponding PRMs location. Identification of the PRB content is necessary to perform this step. In this context, the configuration data is arranged into a set of frames (Xilinx, 2012a), which represent the smallest addressable element in the FPGA configuration memory. For the Virtex-5 FPGA family, used in this work, the configuration frames consist of forty one 32-bit words. A PRB is composed of the following elements: header, commands and configuration data. The header is an opening sequence containing dummy and synchronization words. The commands permit to configure the configuration registers to read or write mode. The configuration data in which we are interested for the bitstreams relocation, are those concerning the FAR and the cyclic redundancy check (CRC) register. The frame address register holds the current frame address and the CRC register provides the data input integrity.

The 32-bits of the FAR are divided into five parts: block type (3 bits), top/bottom (1 bit), row address (5 bits), major column (8 bits), and minor address (7 bits). The Virtex-5 FPGA are divided vertically into two top and bottom parts, where the frames in the top half represent a symmetric images of the frames in the bottom half. Block type: specify the type of resources used in the FPGA like CLB, DSP and BRAM. The FPGA is also divided vertically according to the horizontal clock regions (HCLK). For the top half, the HCLK, regions are numbered from 0 in the middle of the FPGA and the address is upwardly incremented. The bottom part has the same addressing mode, starting with 0 in the middle and downwardly incremented. Each block type has its own column address (or major address), which starts from 0 in the left side. Each column of a block type is composed of a set of frames, which are identified using the minor address. This address varies with the block type (Xilinx, 2012a): for a given column there are 36 minor frames for CLBs, 26 for DSPs, 30 for BRAM, 54 for IOBs, and 4 for GCLK. In the bitstream modification step, the FARs corresponding to initial design are identified and then changed in order to be adapted for each new location (partition). Figure 7 illustrates an example of the FAR modification.

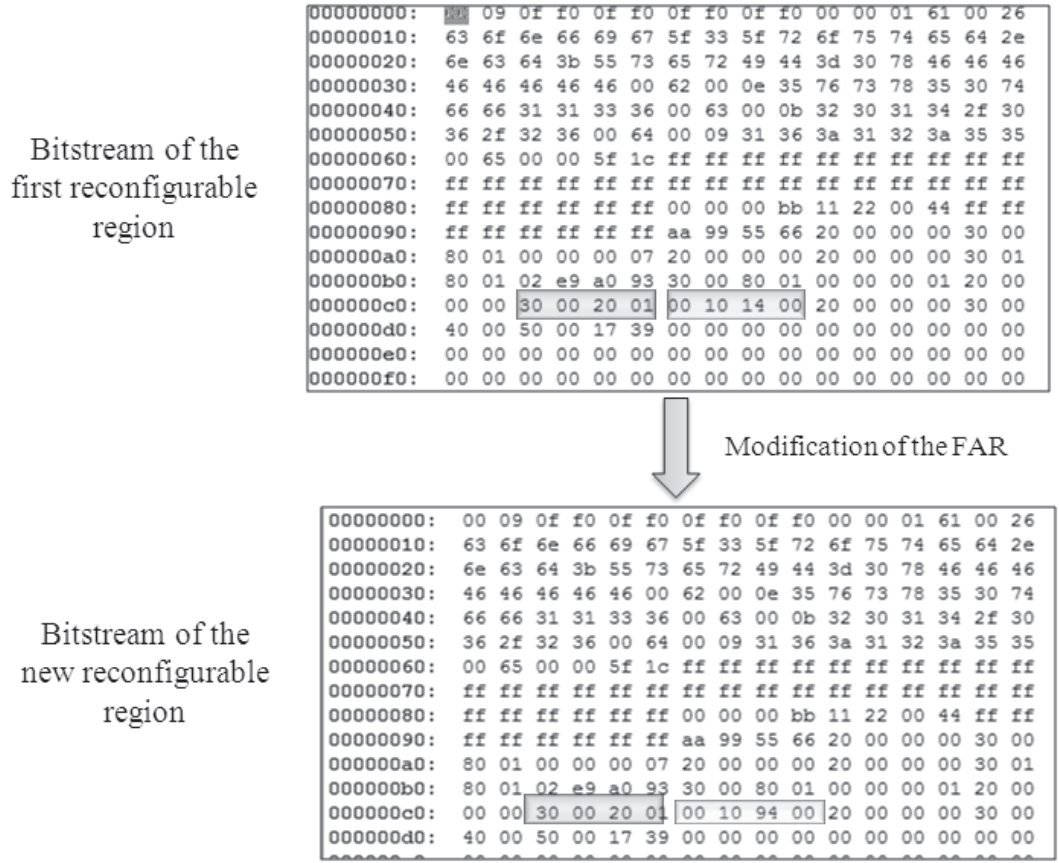

**Fig. 7.** FAR Modification example

## 5. Implementation and experimental results

The proposed relocation method was applied for the video cut detection application, which was implemented using the XUP-ML506 board with a Virtex-5 FPGA device. In our design, we have initially defined 3 PRRs and the rest of the device (the static part) was also defined as reconfigurable, but with only one configuration. Each PRR is partitioned into a number of PRPs to allow the implementation of the PR modules (cut detectors based on CSD with different quantization levels) in more than one possible location. Figure 8 shows the different PRRs and the defined partitions relatively to each module. The hardware module of cut detection based on the CSD with 32 quantization levels(CSD_32) can be placed in two locations, the one based on the CSD with 16 quantization levels (CSD_16) can be located in 5 different locations and the one based on the CSD with 8 quantization levels (CSD_8)can be located in 8 different reconfigurable partitions.

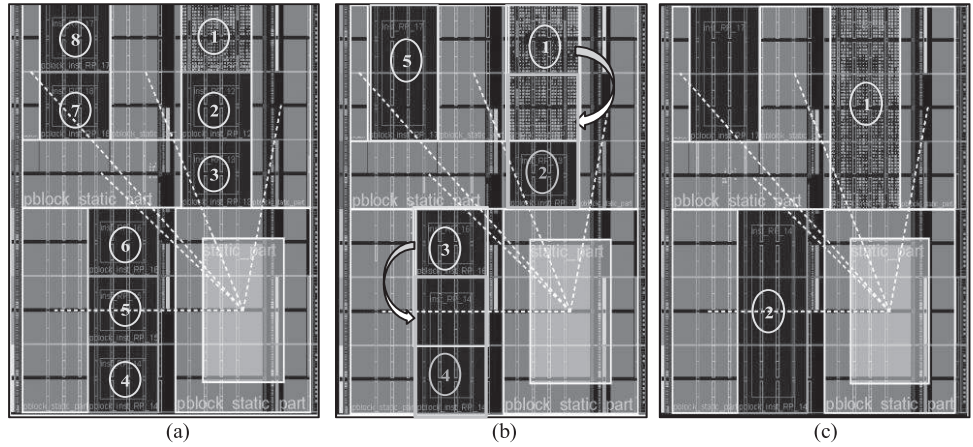

**Fig. 8.** Different possible locations of the CUT detector: a: Cut detector based on the CSD_8, b: Cut detector based on the CSD_16, c: Cut detector based on the CSD_32

Table 3 gives the number and size of the generated bitstreams for different quantization levels (8, 16 and 32) for the design with and without relocation. By applying the design technique based on the classic Xilinx design flow, generating multiple bitstreams for the same PRMis needed: 8 bitstreams for the cut detection module based on the CSD_8, 5 bitstreams for the CSD_16 and 2 bitstreams for the CSD_32. By using the relocation technique, only one bitstream is generated for the initial location of each PRM. The obtained results show that the size of the memory used to store PRBs decreases up to 87.5%, 80% and 50% respectively for the CSD-8, CSD-16 and CSD-32 versions of the CSD.

**Table 3.** Memory optimization results

| CSD version | Without relocation | | With relocation | | Memory optimization |
|---|---|---|---|---|---|
| | BitstreamNumber | Size (Kbyte) | Bitstream Number | Size (Kbyte) | |
| CSD_8 | 8 | 8*112= 896 | 1 | 112 | 87,5% |
| CSD_16 | 5 | 5*224=1120 | 1 | 224 | 80% |
| CSD_32 | 2 | 2*336=672 | 1 | 336 | 50% |

In a Virtex-5 FPGA, a CLB frame is equivalent to 160 LUTs and 40 slices, a BRAM frame consists of 4 BRAM36 memory blocks and a DSP frame is composed of 8 DSP48E blocks arranged vertically. Table 4 provides information on the equivalent number of frames required to implement the different CSD IP versions on a virtex-5 FPGA.

**Table 4.** Frame occupation for the different CSD versions

| CSD IP | Frames | |
|---|---|---|
| | CLB | BRAM |
| **CSD_8** | 9 | 1 |
| **CSD_16** | 16 | 1 |
| **CSD_32** | 27 | 1 |

A main advantage of our method is the adaptation of the PRR to the hardware tasks size. This is based on dividing the PRR into partitions as previously explained. Figure 9 shows the two implementation modes of the multi-quantization levels cut detector for a given PRR: with and without adaptation. The first implementation of the CSD architectures is based on the classic design methods, which do not allow adaptation of the hardware IP cores size to the size of reconfigurable regions (Table 5 without adaptation).In this case, the whole PRR is reserved to implement the three (for region1, 2) or the two (for region 3) versions of CSD based cut detector. This implementation mode leads to inefficient resources utilization.Unused hardware resources of reconfigurable region are wasted and cannot be used for other hardware IP cores implementation.

The second implementation is based on the proposed design method, which allows the adaptation of reconfigurable region to the hardware IP cores size (Table 5 with adaptation). In this implementation mode, the PRRs are partitioned into different partitions adapted to the CSD versions size. The three considered PRRs (PRR1, PRR2, and PRR3) of Figure 9 and their partitioning are defined as follows:

- PRR1 and PRR2 have the same size : 33 frames (27 CLB frames + 6 BRAM Frames)

- PRR3 size = 22 Frames (18 CLB frames + 4 BRAM Frames)
- PRR1 an PRR2 are partitioned into three equal partitions
- PRR3 is partitioned into two equal partitions

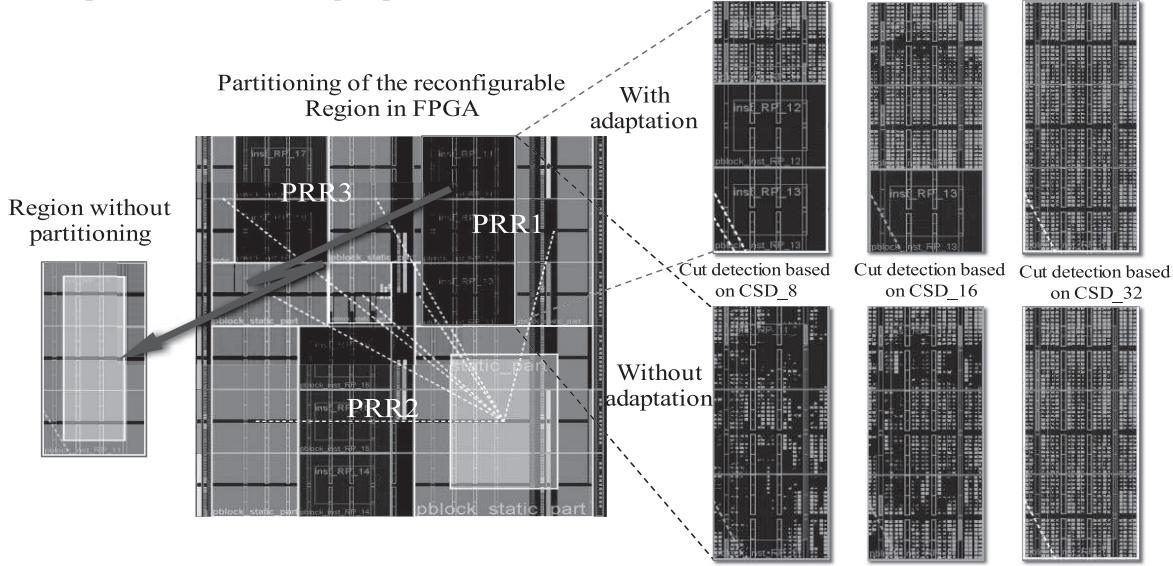

**Fig. 9.** Relocation of PRBs with adaptation

Table 5 gives the hardware resources utilization rates relative to each reconfigurable resource type (Rai, i = CLB or DSP or BRAM) and the global utilization rate (RaT) for the different PRRs (12/ and 3). These rates are calculated for each CSD version. TheRaTvalue also gives information about the wastage rate (Wa), which represents the rate of unused area. The applied metrics (utilization and wastage rates) are defined as follows:

$$Ra_i = \frac{W_i}{N_i}, \quad i \in CLB, DSP, BRAM \quad (5)$$

$$Ra_T = \frac{A_{req}}{A_{res}} = \frac{\sum_i W_i}{\sum_i N_i}, \quad i \in CLB, DSP, BRAM \quad (6)$$

$$Wa = \frac{A_{res} - A_{req}}{A_{res}} = 1 - Ra_T \quad (7)$$

where Areq is the required area, Ares is the reserved area, Wi is the frames number of type i required to implement a given module and Ni is the frames number of type i relative to the reserved partition.

As shown in Table5, the efficiency of the partitions hardware resources can attain a high utilization rates compared to those obtained for the implementation without adaptation. For example, 90 %, 77%, and 84 % rates are obtained for region 1.

**Table 5.** Hardware resources occupation rates

| CSD | Without Adaptation | | | With adaptation | | | |
|---|---|---|---|---|---|---|---|
| | $Ra_{CLB}$ | $Ra_{BRAM}$ | $Ra_T$ | $Ra_{CLB}$ | $Ra_{BRAM}$ | $Ra_T$ | Free Partitions |
| **PRR1/2** | | | | | | | |
| CSD-8 | 33% | 17% | 30% | 82% | 50% | 90% | 2 |
| CSD-16 | 59% | 17% | 51% | 89% | 25% | 77% | 1 |
| CSD-32 | 100% | 17% | 84% | 100% | 17% | 84% | 0 |
| **PRR 3** | | | | | | | |
| CSD-8 | 50% | 25% | 45% | 82% | 50% | 90% | 1 |
| CSD-16 | 89% | 25% | 77% | 89% | 25% | 77% | 0 |

Another benefit of our method is the reconfiguration time optimization. As shown in Table 6, an important gain in terms of reconfiguration time is obtained, when the PRR adaptation is applied. In fact, the total reconfiguration time depends on the size of bitsreams, which itself depends on the PRR size. Without adaptation, the reconfiguration time is the same for the different CSD versions (CSD_8, CSD_16 and CSD_32), because the same reconfigurable area is used to implement the three corresponding hardware modules. However, for the implementation with area adaptation, each module has its own reconfiguration time, which depends on the area of the partition occupied by this module. The obtained results show that the reconfiguration time for the smallest hardware task (cut detection based on the CSD_8) is divided by three, whereas the one corresponding to the cut detection module based on the CSD_16 presents a decrease of 33%.

Table 6. Reconfiguration time evaluation

| n | Reconfiguration times (ms) | | Gain |
|---|---|---|---|
| | Without adaptation | With adaptation | |
| 8 | 0,84 | 0,28 | 66% |
| 16 | 0,84 | 0,56 | 33% |
| 32 | 0,84 | 0,84 | 0% |

The configuration times, shown in Table 6, are calculated using the Equation 8. For an operating frequency of 100 MHz of the configuration access port (ICAP) and a data bus width of 32 bits, a maximum through put of 400KB/ms is reached.

$$Configuration\_time = \frac{bitsream\_size}{configuration\_clock\_frequency \times data\_bus\_widh} \quad (8)$$

## 6. Conclusion

In this paper, we have proposed, dynamic implementation of a video cut detection system based on the MPEG-7 CSD. The proposed design was implemented for 32, 16 and 8 quantization levels. The obtained modules can be loaded depending on the application requirements and located in different possible PRRs using the partial dynamic reconfiguration technique. Our objective was also to generate only one bitstream for each PRM and to use the relocation technique to adapt them to the desired location. The effectiveness of the relocation method, the ability to support variable seize modules, the area occupation efficiency and the reconfiguration time minimization were the key points to be taken into consideration.

A new method for PRB relocation supporting the relocation of variable size hardware tasks was proposed. It allows relocating multiple modules with different resources occupation into different PRRs, while minimizing the wasted hardware resources area. The proposed relocation procedure takes into account the communication between the different PRRs by fixing the partitions pins and the crossing wires routing for all PRRs. The method is also based on partitioning the RRs in a set of partitions adapted to the hardware modules size for an efficient management of the reconfigurable area and the reconfiguration time. In summary, the proposed relocation technique offers many profits that can be very beneficial for our application or any other one: possibility of executing the module generated for a specific PRR on different possible locations, reducing the number of PRBs stored in memory for the same task, increasing the area occupation efficiency and reducing reconfiguration time.

All the issues handled in this work, have been experimentally validated throughout the proposed application (video CUT detection based on the CSD). A Virtex 5 FPGA was used for the system implementation. Only one bitstream was generated for the three versions of the developed module (for 32, 16 and 8 quantization levels). These bitstreams can be modified in the reconfiguration phase to be adapted to a new location and successfully loaded to the FPGA, without any communication (partitions pins and routing paths) problem. The experimental results show decrease of the amount of memory used to store PRBs, the significant gain in terms of hardware resources utilization efficiency and reconfiguration time reduction compared to standard Xilinx DPR design flow.

## References

**Aljazzaf, Z. M. (2015).** Modelling and measuring the quality of online services. Kuwait Journal of Science, **42**(3):134–157.

**Becker, T., Luk, W. & Cheung, P. Y. K. (2007).** Enhancing relocatability of partial bitstreams for run-time reconfiguration. Proceedings 2007 IEEE Symposium on Field-Programme Custom Computing Machines.

**Beckhoff, C., Koch, D. & Torresen, J. (2012).** GoAhead: A partial reconfiguration framework. Proceedings of the 2012 IEEE 20th International Symposium on Field-Programmable Custom Computing Machines.

**Ben Abdelali, A., Hannachi, M., Touil, L. & Mtibaa, A. (2014).** Adequation and hardware implementation of the color structure descriptor for real-time temporal video segmentation. Journal of Real-Time Image, 1–20.

**Ben Abdelali, A. & Mtibaa, A. (2005).** Toward hardware implementation of the compact color descriptor for real time video indexing. Advances in Engineering Software, **36**(7):475–486.

**Cernekova, Z., Pitas, I. & Nikou, C. (2006).** Information theory-based shot cut/fade detection and video summarization. IEEE Transactions on Circuits and Systems for Video Technology, **16**(1):82–91.

**Corbetta, S., Morandi, M., Novati, M., Santambrogio, M. D., Sciuto, D. & Spoletini, P. (2009).** Internal and external bitstream relocation for partial dynamic reconfiguration. IEEE Transactions on Very Large Scale Integration (VLSI) Systems, **17**(11):1650–1654.

**Drahonovsky, T., Rozkovec, M. & Novak, O. (2013).** Relocation of reconfigurable modules on Xilinx FPGA. Proceedings of the 2013 IEEE 16th International Symposium on Design and Diagnostics of Electronic Circuits and Systems, DDECS 2013, 175–180.

**Flynn, A., Gordon-Ross, A. & George, A. D. (2009).** Bitstream relocation with local clock domains for partially reconfigurable FPGAs. In 2009 Design, Automation & Test in Europe Conference & Exhibition.

**Hannachi, M., Rabah, H., Jovanovic, S., Abdelali, A. & Mtibaa, A. (2014).** Efficient relocation of variable-sized hardware tasks for FPGA-based adaptive systems. In Microelectronics (ICM), 2014 26th International Conference on Doha, Qatar.

**Haroldsen, T., Nelson, B. & Hutchings, B. (2015).** RapidSmith 2: A framework for BEL-level CAD exploration on Xilinx FPGAs. In Proceedings of the 2015 ACM/SIGDA International Symposium on Field-Programmable Gate Arrays (bll 66–69) New York, NY, USA: ACM.

**Hong, C., Benkrid, K., Isa, N. & Iturbe, X. (2014).** A run-time reconfigurable system for adaptive high performance efficient computing. ACM SIGARCH Computer Architecture News, **41**(5):113–118.

**Horta, E. L., Lockwood, J. W. & Louis, S. (2001).** PARBIT : A tool to transform bitfiles to implement partial reconfiguration of field programmable gate arrays ( FPGAs )

**Ichinomiya, Y., Usagawa, S., Amagasaki, M., Iida, M., Kuga, M. & Sueyoshi, T. (2012).** Designing flexible reconfigurable regions to relocate partial bitstreams. Proceedings of the 2012 IEEE 20th International Symposium on Field-Programmable Custom Computing Machines, FCCM 2012, 241.

**Kalte, H., Lee, G., Porrmann, M. & Rückert, U. (2005).** REPLICA: A bitstream manipulation filter for module relocation in partial reconfigurable systems. In Proceedings - 19th IEEE International Parallel and Distributed Processing Symposium, IPDPS 2005 (Vol 2005)

**Koch, D., Beckhoff, C. & Teich, J. (2008).** Recobus-builder - a novel tool and technique to build statically and dynamically reconfigurable systems for FPGAs. Proceedings - 2008 International Conference on Field Programmable Logic and Applications, FPL, 119–124.

**Krulikovská, L. & Polec, J. (2012).** An efficient method of shot cut detection. International Journal of Electrical, Computer, Energetic, Electronic and Communication Engineering, **6**(1):356–360.

**Lysaght, P., Blodget, B., Mason, J., Young, J. & Bridgford, B. (2006).** Invited paper: enhanced architectures, design methodologies and CAD tools for dynamic reconfiguration of Xilinx FPGAs. 2006 International Conference on Field Programmable Logic and Applications, 1–6.

**Manjunath, B. S., Ohm, J. R., Vasudevan, V. V & Yamada, A. (2001).** Color and texture descriptors. IEEE Transactions on Circuits and Systems for Video Technology, **11**(6):703–715.

**Marques, N., Rabah, H., Dabellani, E. & Weber, S. (2014).** A novel framework for the design of adaptable reconfigurable partitions for the placement of variable-sized IP cores. IEEE Embedded Systems Letters, 0663(c), 45-48.

**Sohanghpurwala, A. A., Athanas, P., Frangieh, T. & Wood, A. (2011).** OpenPR: An open-source partial-reconfiguration toolkit for xilinx FPGAs. IEEE International Symposium on Parallel and Distributed Processing Workshops and Phd Forum, (Xdl), 228–235.

**Sudarsanam, A., Kallam, R. & Dasu, A. (2010).** PRR-PRR dynamic relocation. IEEE Computer Architecture Letters, **8**(2):44–47.

**Touiza, M., Ochoa-Ruiz, G., Bourennane, E. B., Guessoum, A. & Messaoudi, K. (2013).** A novel methodology for accelerating bitstream relocation in partially reconfigurable systems. Microprocessors and Microsystems, **37**(3):358–372.

**Vertan, C., Zamfir, M. & Dr^, A. (2011).** Mpeg-7 compliant generalized structure descriptor for still image indexing. Computer, 2481–2484.

**Vipin, K. & Fahmy, S. A. (2015).** Mapping adaptive hardware systems with partial reconfiguration using CoPR for Zynq. 2015 NASA/ESA Conference on Adaptive Hardware and Systems (AHS), 1–8.

**Xilinx. (2012a).** Virtex-5 FPGA Configuration User Guide (Vol 191)

**Xilinx. (2012b).** Xilinx Partial Reconfiguration User Guide, UG702, 1–124.

**Yu, J. & Srinath, M. D. (2001).** An efficient method for scene cut detection. Pattern Recognition Letters, **22**:1379–1391.

# طريقة فعالة لإدارة مناطق قابلة لإعادة التشكيل للأنظمة القائمة على مصفوفة البوابات المنطقية القابلة للبرمجة FPGA متكيفة وديناميكية

مروى الحناشي[2]،[1]،[*]، عبد السلام عبد العالي[1]، حسن رباح[2]، عبد اللطيف مطيبع[1]

[1]مختبر الإلكترونيات والإلكترونيات الدقيقة، جامعة المونستير، تونس

[2]معهد جين لامور (IJL) UMR7198، جامعة لورين، فرنسا

*marwa.hannachi@univ_lorraine.f

## خـلاصـة

يمكن أن تستفيد الأنظمة المرنة القائمة على هيكلية مصفوفة البوابات المنطقية القابلة للبرمجة (FPGA) بدرجة كبيرة من درجة المرونة العالية التي تتيحها عملية إعادة التشكيل الجزئي الديناميكي (DPR). فباستخدام هذه التقنية، يمكن تحميل مهام الأجهزة وإعادة تحميلها عدة مرات عند الطلب وفقاً لمتطلبات النظام. وفي هذا البحث، نقترح استخدام التشكيل الجزئي الديناميكي (DPR) للتنفيذ الديناميكي والمرن لبرنامج يكشف عن القطع في تسجيل فيديو اعتماداً على واصفات البنية اللونية (CSD) لمصفوفة البوابات المنطقية القابلة للبرمجة 7 (FPEG-7). تم اختبار سيناريوهات مختلفة في التنفيذ المقترح. واعتماداً على البرنامج التطبيقي ومتطلبات النظام، يمكن تحميل وحدة برمجية لواصفات البنية اللونية (CSD) في أي وقت بحجم متغير (وفقاً لنسخة مختلفة من CSD) وتخصيصها في مختلف المناطق القابلة لإعادة التشكيل المحتملة. وينطوي هذا التنفيذ على العديد من المشاكل المتصلة بالإدارة، والنقل، وإدارة المنطقة القابلة لإعادة التشكيل. وسنوضح كيف نجحنا في هذا التنفيذ من خلال استخدام طريقة التصميم المناسبة. وتم اقتراح هذه الطريقة لدعم إدارة مهام الأجهزة ذات الأحجام المتغيرة على إعادة التشكيل الجزئي الديناميكي DPR في الأنظمة المرنة القائمة على مصفوفات البوابات المنطقية القابلة للبرمجة FPGA. وهو يسمح بالتعامل بكفاءة مع منطقة قابلة لإعادة التشكيل ونقل الوحدات البرمجية القابلة لإعادة التشكيل في مختلف المناطق المحتملة. وتُظهر نتائج التنفيذ للبرنامج التطبيقي موضع الدراسة تحسناً هاماً من حيث وقت التشكيل (حتى 66%) وتخزين الذاكرة (حتى 87%) ومعدل استخدام موارد الأجهزة بكفاءة (حتى 90%).